\documentclass[aps,prl,twocolumn,showpacs]{revtex4}

\usepackage{graphicx}

\newcommand{\citeS}[1]{ \cite{#1}}
\newcommand{\citeR}[1]{\cite{#1}}

\begin{document}
\title{Coherent Optical Memory with High Storage Efficiency and Large Fractional Delay}
\author{Yi-Hsin Chen,$^{1}$ Meng-Jung Lee,$^{1}$ I-Chung Wang,$^{1}$ Shengwang Du,$^{2}$ Yong-Fan Chen,$^{3}$ Ying-Cheng Chen,$^{4,1}$ and Ite A. Yu$^{1}$}
\email{yu@phys.nthu.edu.tw}
\affiliation{
$^{1}$Department of Physics and Frontier Research Center on Fundamental and Applied Sciences of Matters, National Tsing Hua University, Hsinchu 30013, Taiwan \\ 
$^{2}$Department of Physics, The Hong Kong University of Science and Technology, Clear Water Bay, Kowloon, Hong Kong, China \\ 
$^{3}$Department of Physics, National Cheng Kung University, Tainan 70101, Taiwan \\ 
$^{4}$Institute of Atomic and Molecular Sciences, Academia Sinica, Taipei 10617, Taiwan}
\begin{abstract}
A high-storage efficiency and long-lived quantum memory for photons is an essential component in long-distance quantum communication and optical quantum computation. Here, we report a 78\% storage efficiency of light pulses in a cold atomic medium based on the effect of electromagnetically induced transparency (EIT). At 50\% storage efficiency, we obtain a fractional delay of 74, which is the best up-to-date record. The classical fidelity of the recalled pulse is better than 90\% and nearly independent of the storage time, as confirmed by the direct measurement of phase evolution of the output light pulse with a beat-note interferometer. Such excellent phase coherence between the stored and recalled light pulses suggests that the current result may be readily applied to single photon wave packets. Our work significantly advances the technology of EIT-based optical memory and may find practical applications in long-distance quantum communication and optical quantum computation.
\end{abstract}
\maketitle

\newcommand{\FigOne}{
	\begin{figure}[t] 
	\includegraphics[width=7cm]{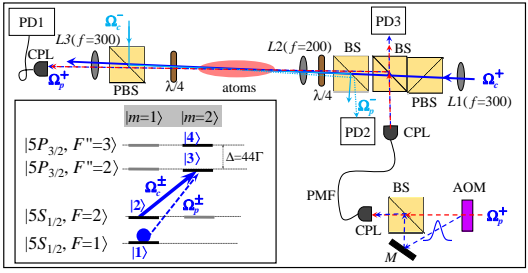}
	\caption{
Schematic experimental setup. PD1, PD2, and PD3: photo-detectors; PMF: polarization-maintained optical fiber; CPL: optical fiber coupler; BS: cubed beam splitter; PBS: cubed polarizing beam splitter; $\lambda/4$: zero-order quarter-wave plate; $L$1-$L$3: lenses with the focal length $f$ specified in units of mm. $M$: mirror; AOM: 80-MHz acoustic optical modulator. Inset: relevant energy levels and laser excitations in the experiment.}
	\label{fig:setup}
	\end{figure}
}
\newcommand{\FigTwo}{
	\begin{figure}[t] 
	\includegraphics[width=7cm]{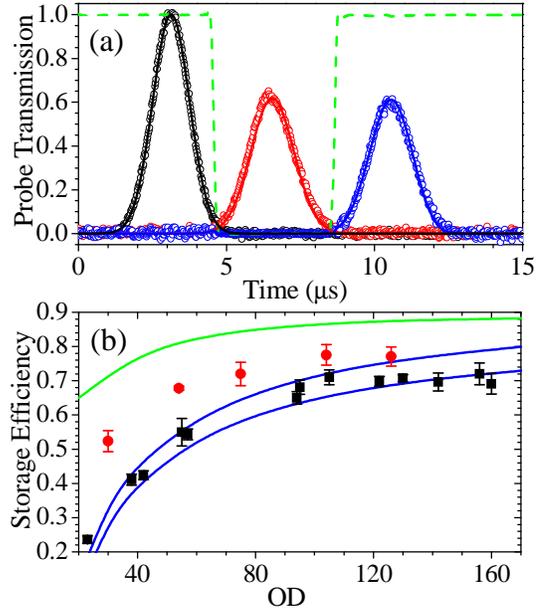}
	\caption{
(a) Experimental data of the input, slow-light output, and stored and then recalled pulses in the forward configuration are shown by black, red, and blue open circles, respectively. Solid lines represent the input pulse used in the calculation and the theoretically predicted outputs. The coupling field employed in the light-storage measurement is shown by the dashed line. OD = 156, $\Omega_c^+$ = 1.1$\Gamma$, $\gamma$ (the decoherence rate during pulse propagation) = 5.5$\times$$10^{-4}$$\Gamma$, and $\tau_{\rm coh}$ (the coherence time during light storage) = 98 $\mu$s. The storage efficiency (SE) determined from the measurement is 69\%. (b) Experimental data and theoretical predictions of SEs as functions of OD are shown in the forward (black squares and blue lines) and backward (red circles and green lines) retrievals. In the forward retrieval, $\Omega_c^+$ = 0.3$\sim$1.1$\Gamma$ and increases with OD; $\gamma$ = 2.0$\times$$10^{-4}$$\Gamma$ (upper blue line) or $5.5\times10^{-4} \Gamma$ (lower blue line). In the backward retrieval, $\Omega_c^{\pm} =$ 1.2$\Gamma$ and $\gamma$ = 5.5$\times$$10^{-4}$$\Gamma$.}
	\label{fig:forward}
	\end{figure}
}
\newcommand{\FigThree}{
	\begin{figure}[t] 
	\includegraphics[width=7cm]{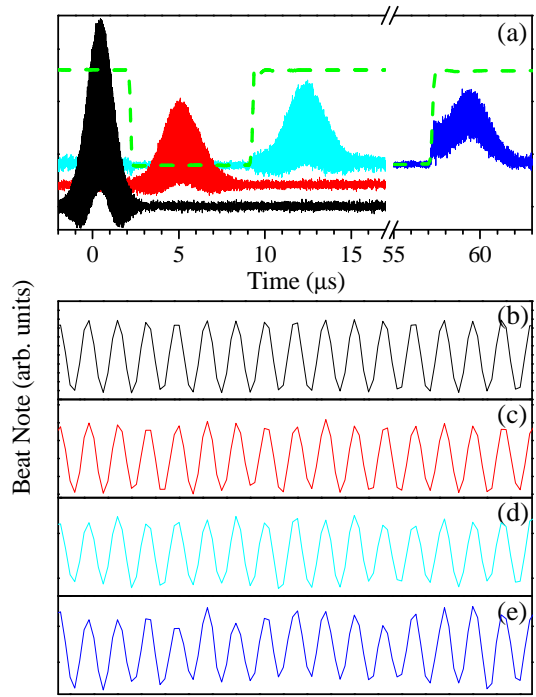}
	\caption{
(a) Beat notes of the input (black) and slow-light output (red) pluses and the recalled pulses after 7 $\mu$s (cyan) and 55 $\mu$s (blue) storage times. Green dashed lines are the coupling field employed in the light-storage measurement. (b)-(e) Zooming in on the time intervals around the pulse peaks with the same horizontal span of 200 ns. The ratio of the vertical spans in (b)-(e) is 4:2:2:1. The frequency of beat notes is 80 MHz.}
	\label{fig:beat}
	\end{figure}
}
\newcommand{\FigFour}{
	\begin{figure}[t] 
	\includegraphics[width=7cm]{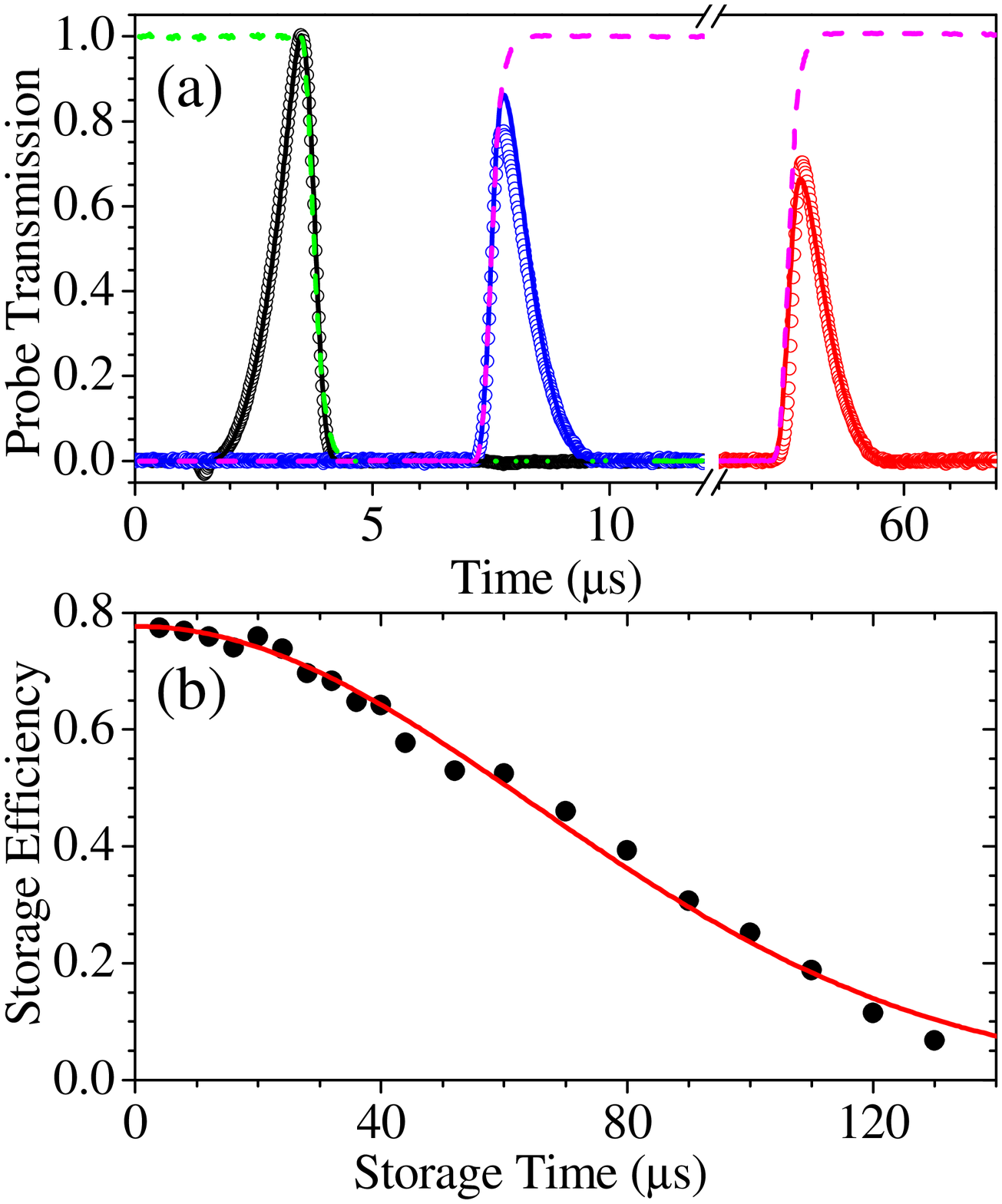}\label{fig:fig4}
	\caption{
(a) Based on the time-space-reversing method, experimental data and the theoretical predictions of the input pulse and the stored and then backward recalled pulses after 4 $\mu$s and 54 $\mu$s storage times. Blue and magenta dashed lines are the forward and backward coupling fields and other legends are the same as those in Fig.~2(a). The input pulse shape is optimized for the maximum SE. OD = 104, $\Omega_c^+$ = $\Omega_c^-$ = 1.2$\Gamma$, $\gamma$ = 5.5$\times$$10^{-4}$$\Gamma$, and $\tau_{\rm coh}$ = 104 $\mu$s. The storage efficiencies (SE) determined from the measurements are 78\% and 56\% at 4 $\mu$s and 54 $\mu$s storage time, respectively. The FWHM of the input pulse is about 0.88 $\mu$s. (b) SE as a function of storage time. The solid line is the best fit with the Gaussian-decay function, $\exp(-t^2/\tau_{\rm coh}^2)$, which determines $\tau_{\rm coh}$ is 92 $\mu$s.}
	\label{fig:backward}
	\end{figure}
}

Quantum memory\citeS{QuantumEIT1,QuantumEIT2,QuantumEIT3,QuantumGEM1,QuantumGEM2,ClassicalGEM,FidelityGEM,ClassicalRaman} is essential for quantum information processing, including quantum communication\citeS{communication1,communication2,Cirac} and quantum computation\citeS{computation1}. Using quantum repeaters will be a practical protocol for implementing long-distance quantum communication without suffering from transmission loss\citeS{DLCZ,repeater1,repeater3}. The scheme proposed in Ref.~\citeR{DLCZ} divides a long distance into shorter elementary channels, stores and retrieves entanglement pairs, and extends the transmission distance via entanglement swapping. Quantum memory, a storage device mapping quantum state between light and matter, is a crucial component of the quantum repeater. Processing a particular task or waiting for the completion of others requires a quantum state to be stored in memory for a long enough time. Therefore, quantum memory with high storage efficiency (SE), which is defined as the ratio of recalled to input photon energies, and long coherence time are the keys to successful operation of long-distance quantum communication and quantum information processing.

The fractional delay (FD) or delay-bandwidth product at 50\% SE is a possible figure of merit for a memory in the no-cloning limit\citeS{Tolerance1,Tolerance2} or in the one-way quantum computation\citeS{Tolerance3}, where FD is defined as the ratio of storage time to the full-width-half-maximum (FWHM) pulse duration. Several memory devices based on different mechanisms, such as gradient photon echo, Raman interaction, and electromagnetically induced transparency (EIT), have been proposed and experimentally demonstrated. With the gradient echo memory, a maximum SE of 87\% as well as the FD of 11 at 50\% SE for classical light was demonstrated \citeR{ClassicalGEM}, and the recall fidelity can be as high as 98\% for coherent pulses containing around one photon \cite{FidelityGEM}. With a far-off-resonant Raman transition, Ref.~\citeR{ClassicalRaman} showed Raman memory with an SE of 43\% and a coherence time of approximately 1 $\mu$s for 300-ps coherent light pulses. 

Slowing and storing light using the EIT effect\citeS{HarrisEIT,DSP} has been intensively explored in the past two decades owing to its potential applications in optical information processing and optical communication, as well as its compatibility with quantum state operations. Previous theory predicted that the SE of the EIT-based memory could approach unity in an atomic ensemble at a high optical density (OD)\citeS{Reversing}. However, experiments in hot vapor cells have revealed that a maximum SE is only about 43\% at the optimum OD and further increasing OD even reduces the SE\citeS{ReversingExpt,FWMonEIT1,FWMonEIT2}. Recent demonstrations with cold atoms at high OD also showed a saturation value of SE at 50\% for both coherent pulses\citeS{Du1} and single-photon wave packets\citeS{Du2}. These previous experiments seemingly excluded the EIT medium as an efficient quantum memory. In this paper, we report a high-performance EIT-based memory with an SE of 78\% and a coherence time of 98$\pm$6 $\mu$s in a cold atomic medium. This work is the first experimental demonstration that the time-space-reversing method proposed in Ref.~\citeR{Reversing} can enhance the SE of the EIT-based memory. The FD at 50\% SE is 74$\pm$5, which, to our knowledge, is the best record to date. We also demonstrated that the classical fidelity of the recall pulse is better than 90\% as measured by a beat-note interferometer\citeS{Beatnote}. Our results suggest that EIT-based quantum memory can be used for practical quantum information applications.

The experiment was carried out in a cigar-shaped cloud of cold $^{87}$Rb atoms produced by a magneto-optical trap (MOT)\citeS{CigarMotYu}. Typically, we trapped $1.0\times10^9$ atoms with a temperature of about 300 $\mu$K\citeS{AtomTempYu}. The experimental setup is shown in Fig.~1. We pumped the whole population to a single Zeeman ground state $|1\rangle$ as shown in the inset of Fig.~1. The probe (Rabi frequency of which is denoted as $\Omega_p$) and coupling fields all had $\sigma_+$ polarization. They drove the $|1\rangle$$\rightarrow$$|3\rangle$ and $|2\rangle$$\rightarrow$$|3\rangle$ transitions, respectively, forming the $\Lambda$ configuration of EIT. The spontaneous decay rate $\Gamma$ of the excited state $|3\rangle$ is $2\pi\times 6$ MHz. In this study, the stored probe pulse can be retrieved by the coupling field $\Omega_c^+$ (or $\Omega_c^-$) in a forward (or backward) direction. The probe pulse was focused to a beam size with an $e^{-2}$ full width of 150 $\mu$m and propagated along the major axis of the atom cloud, denoted as $\theta = 0^{\circ}$. The forward and backward coupling fields were collimated with beam diameters of 5 and 3 mm, propagating along the directions of $\theta = 0.4^{\circ}$ and $180.4^{\circ}$. We set the peak power of the input probe pulse to about 100 nW. The collection efficiencies of the photo detectors for the forward and backward probe pulses were 52\% and 68\%, respectively. Before performing each measurement, we employed the dark and compressed MOT for about 7 ms to increase the atomic density and the OD of the system. Other details of the experimental setup can be found in Ref.~\citeR{AOSbySLP}.

\FigOne

\FigTwo

We first worked with the forward-retrieval configuration. The experimental data of the input, slow-light output, and stored and then recalled pulses are shown by the open circles in Fig.~2(a). The temporal Gaussian profile of the input probe pulse had an FWHM of 1.5 $\mu$s. The solid lines in the figure are the theoretical predictions. In the calculation, we numerically solved the Maxwell-Schr\"{o}dinger equation of the light pulse and the optical Bloch equation of the atomic density-matrix operator\citeS{SlpYu09,SM}. The good agreement between the data and the predictions indicates that the measurements were taken under OD = 156 and $\Omega_c^+$ = 1.1$\Gamma$. We directly integrated the areas below the input and output probe pulses and used their ratio to determine the SE. As shown in Fig.~2(a), the SE of a 4-$\mu$s storage time was about 69\% which already exceeded previous results for EIT-based storage. Some studies suggest that an unwanted four-wave mixing (FWM) process degrades the SE and limits the maximum achievable SE to about 43\% with hot atomic vapors\citeS{FWMonEIT1,FWMonEIT2}. Here, the experimental data with the cold atoms agree with the theoretical predictions calculated without the FWM process. We also found theoretically that the SEs with and without the FWM process differ by less than 0.2\%. Thus, the FWM effect can be negligible in our system and cold atomic samples. All the equations used for the theoretical predictions and their details are described in Appendix.

The measured SE as a function of OD is shown by the black squares in Fig.~2(b). For each OD, the width of the Gaussian probe pulse was the same and we optimized the coupling Rabi frequency $\Omega_c^+$ to maximize the SE. Because the EIT bandwidth is proportional to $(\Omega_c^+)^2/\sqrt{{\rm OD}}$, a higher OD resulted in a larger optimum $\Omega_c^+$ as expected. The SE data initially increased with the OD but eventually became saturated. This saturation phenomenon made us realize the coupling field also drove the $|2\rangle$$\rightarrow$$|4\rangle$ transition with a detuning of 44$\Gamma$, as shown in Fig.~1. The undesired transition induced the switching effect, destroying the probe pulse\citeS{PS}. A larger Rabi frequency of the $|2\rangle$$\rightarrow$$|4\rangle$ transition or a longer probe pulse made the switching more noticeable. The switching effect is equivalent to increasing the relaxation rate of the ground-state coherence, $\gamma$. In Fig.~2(b), the blue lines are the theoretical predictions of two different $\gamma$'s. At lower (higher) ODs, the SE data points are closer to the predictions of the smaller (larger) $\gamma$ as expected, because the $\Omega_c^+$ used in the measurement was smaller (larger). This explains the observed saturation phenomenon of SE. Please note that during the storage the coupling field was turned off so that the switching effect did not exist. Hence, the coherence time during the storage, $\tau_{\rm coh}$, could not be directly derived from $\gamma$. The measurement of $\tau_{\rm coh}$ will be shown later.

Memory must possess high fidelity for practical applications. The classical fidelity is defined by
\begin{equation} \label{fidelity}
	\frac{\left| \int E^{\ast}_{\rm in}(t-t_d) E_{\rm out}(t) dt \right|^2}
		{\left[ \int |E_{\rm in}(t)|^2 dt \right]
		\left[ \int |E_{\rm out}(t)|^2 dt \right]},
\end{equation}
where $E_{\rm in}$ and $E_{\rm out}$ are the electric fields of input and output pulses and $t_d$ is the delay time. Equation~(1) is similar to $|\langle\psi_{\rm in} |\psi_{\rm out}\rangle|^2 / (\langle\psi_{\rm in} |\psi_{\rm in}\rangle \langle\psi_{\rm out} |\psi_{\rm out}\rangle)$, but we replaced the wave function $\psi$ with the electric field here. To obtain the fidelity, we measured the phase evolution of pulses with a beat-note interferometer\citeS{Beatnote}. The pulsed probe field spatially overlapped the continuous-wave reference field, as depicted in the bottom-right corner of Fig.~1. The intensity of the reference field was about one tenth of that of the probe pulse peak. The combined fields produced the beat note at 80 MHz and entered a polarization-maintained fiber (PMF). Coming out of the PMF, the combined fields split into two beams. One beam was directly detected by PD3, the beat note signal of which served as the phase reference. The other propagated through the atoms and was detected by PD1 producing the data shown in Fig.~3. Other details of the beat-note interferometer can be found in Ref.~\citeR{Beatnote}. The high-frequency signal in Fig.~3(a) is the beat note that was averaged over 1436 traces. We can zoom in on different time intervals and clearly see the phase evolution of the beat note as demonstrated in Figs.~3(b)-3(e). The beat note provided the phase as well as amplitude information on the electric field in the output probe pulse. Using Eq.~(1) (the integration range covers the $e^{-2}$ pulse width) and the beat note signals, we determined that the classical fidelity of the recalled pulse after a 7-$\mu$s (or 55-$\mu$s) storage time was 94\% (or 90\%). The theoretical fidelity was 97\% due to the output pulse width being broadened to 1.7 times the input one. The measured fidelity was limited by the signal-to-noise ratio in the beat-note signal. The beat-note amplitude in the data yielding a fidelity of 94\% was about twice that yielding a fidelity of 90\%. We increased the reference field intensity to double the beat note amplitude and the above fidelity was improved to 97\% (or 93\%). Thus, the classical fidelity was influenced very little by storage time. The excellent phase coherence between the stored and recalled light pulses suggests that the current results may be readily applied to single-photon wave packets.

\FigThree

Reference~\citeR{Reversing} proposed a time-space-reversing method for improving the SE with a backward-retrieval configuration. A representative example is shown in Fig.~4(a). The probe pulse was sent to the medium in the forward direction, stored by turning off $\Omega_c^+$, and retrieved in the backward direction by turning on $\Omega_c^-$. The retrieved probe pulse provided the reversed temporal profile for the input probe pulse in the next run. The forward write-in and backward read-back process was run iteratively until the input and output pulse shapes were nearly in the time-reversal symmetry and the SE approached the maximum value asymptotically. The optimized input probe shape had a slowly varying front, with the frequency components well within the EIT bandwidth, and a sharp-edged rear, corresponding high-frequency components. The slowly varying front propagated through the entire medium with minimum loss. The sharp-edged rear was forward stored near the medium entrance and backward retrieved without propagating through the medium, leading to a low energy loss. Representative data at storage times of 4 and 54 $\mu$s are shown by the open circles in Fig.~4(a). The theoretcial predictions shown by the solid lines and the experimental data had nearly the same shape but with slight differences in the amplitudes\citeS{SM}. The measured SE was enhanced to 78\% with the backward retrieval plus optimum pulse shape. This is also the first experimental demonstration confirming the time-space-reversing theory in the EIT system\citeS{Reversing}.

\FigFour

The red circles in Fig.~2(b) show the measured optimal SE as a function of OD in the backward-retrieval configuration. Because the coupling Rabi frequency does not affect the SE and only determines the optimum pulse shape in this method\citeS{ReversingExpt}, we kept $\Omega_c^{\pm}$ approximately the same for all ODs. Each measured datum was smaller than its theoretical prediction. The discrepancy may be due to phase mismatch during the backward retrieval\citeS{phasematch}. A phase mismatch caused by a beam misalignment of 0.1$^{\circ}$ can reduce the retrieved energy by about 4\%. Both the measured and predicted SEs behaved similarly, increasing with the OD and approaching an asymptotic value. Figure~4(b) shows the storage efficiency as a function of storage time. The experimental data behaved like a Gaussian-decay function, i.e., $\exp(-t^2/\tau_{\rm coh}^2)$ where $\tau_{\rm coh}$ is the coherence time, suggesting that the atomic motion was the dominant decoherence mechanism. Best fit of the data gives $\tau_{\rm coh}$. The average of day-to-day $\tau_{\rm coh}$ is 98$\pm$6 $\mu$s. At the 50\% SE, we obtain an excellent fraction delay of 74$\pm$5.

In the backward-retrieval configuration, the input and recall pulses propagated in opposite directions and were detected by two different detectors (PD1 and PD2 in Fig.~1). This makes it extremely difficult to measure their phase correlation. Because the beat-note data of Fig.~3 showed excellent phase correlation between the input and forward-recalled pulses, the phase of the input pulse must have been encoded to the atomic spin wave during the storage. The retrieval process in either the forward or backward direction maps both amplitude and phase of the spin wave to the recall pulse. Therefore, the input and backward-recalled pulses should also be phase coherent. Under an assumption of perfect phase correlation, we took the square root of the input and retrieved pulses (after time reversal) as the fields in Eq.~(1) and estimate the classical fidelities of the two recall data in Fig.~4(a) all to be greater than 99\%, which is better than the theoretical fidelity in the forward retrieval. The superior result is due to nearly perfect similarity between the input and time-reversed output pulse shapes.

In conclusion, we have demonstrated an efficient EIT-based optical memory with cold atoms. Our results in the forward retrieval show a SE of 69\%. Using the space-time-reversing method plus optimum pulse shape, we further enhanced the SE to 78\%. This efficiency is normalized with respect to the collection efficiencies of the photodetectors in the forward and backward directions. The saturation of SE at a high OD is mainly caused by the dephasing rate of about 5.5$\times$$10^{-4}$$\Gamma$. In our system, the saturation resulted from the far-detuned switching effect and the laser frequency instability, rather than from the FWM process which limits the SE with hot atomic vapors. The measured decay time of the memory is about 98 $\mu$s. Allowing the SE to be 50\%, we obtained a large FD of 74. We also employed a beat-note interferometer to verify that the classical fidelity of the recalled pulse was better than 90\% and nearly independent of the storage time. There is excellent phase coherence between input and output pulses as well as good agreement between theoretical and experimental values of fidelity. Although the experiments were done with coherent optical pulses, the EIT light-matter interface may possibly be applied to single-photon quantum states due to its quantum nature. Our work may bring the EIT-based quantum memory toward practical quantum information applications.

\section*{ACKNOWLEDGEMENTS}
This work was supported by the National Science Council of Taiwan under Grants No. 100-2628-M-007-001 and No. 101-2923-M-007-002 and by National Tsing Hua University under Grant No. 101N2713E1. S. D. was supported by the Hong Kong Research Grants Council (Project No. 600710). The Hsinchu team is the partner in EU FP7 IRSES project COLIMA (Contract No. PIRSES-GA-2009-247475). The authors thank Professor P. K. Lam for valuable comments on the manuscript.


\section*{Appendix}
We made the theoretical predictions by numerically solving the Maxwell-Schr\"{o}dinger equation of the light pulse and the optical Bloch equation of the atomic density-matrix operator. The inset in Fig.~1 of the main text shows the three-level EIT system formed by the two ground states of $|1\rangle$ and $|2\rangle$ and the excited state of $|3\rangle$. The probe and coupling fields drive the $|1\rangle \rightarrow |3\rangle$ and $|2\rangle \rightarrow |3\rangle$ transitions with the Rabi frequencies of $\Omega_p$ and $\Omega_c$, respectively. Considering the very weak probe pulse as the perturbation, we used the following equations to calculate the theoretical predictions of slow light and storage and retrieval of light [31]:
\begin{eqnarray}
    \frac{1}{c}\frac{\partial}{\partial t} \Omega_p
        + \frac{\partial}{\partial z} \Omega_p
        =  i \eta \rho_{31}, \\
   \frac{\partial}{\partial t} \rho_{31}
        = \frac{i}{2} \Omega_p +\frac{i}{2} \Omega_c \rho_{21}
        - \frac{\Gamma}{2} \rho_{31}, \\
    \frac{\partial}{\partial t} \rho_{21}
        = \frac{i}{2} (\Omega_c)^* \rho_{31} - \gamma \rho_{21},
\end{eqnarray}
where $\rho_{ij}$'s are the elements of the density-matrix operator; $\Gamma$ is the spontaneous decay rate of the excited state and equal to 2$\pi$$\times$6 MHz in our case; $\eta$ = $\alpha\Gamma/(2L)$; $\alpha$ and $L$ are the optical density (OD) and length of the medium; $\gamma$ is the relaxation rate of the ground-state coherence. 

In the four-wave mixing (FWM) process, the same coupling field in the EIT scheme also drives the far-detuned $|1\rangle \rightarrow |3\rangle$ transition and generates the sum-frequency field ($\Omega_s$) of the $|3\rangle \rightarrow |2\rangle$ transition with the help of $\rho_{21}$. We follow the formulas given by Ref.~[24], which experimentally and theoretically studied this FWM effect, and modify Eqs.~(S1)-(S3) to 
\begin{eqnarray}
    \frac{1}{c}\frac{\partial}{\partial t} \Omega_s
        + \frac{\partial}{\partial z} \Omega_s
        = i\eta (\rho_{21})^* \frac{\Omega'_c / \Gamma}{2\Delta /\Gamma_l}, \\
    \frac{1}{c}\frac{\partial}{\partial t} \Omega_p
        + \frac{\partial}{\partial z} \Omega_p
        = i\eta \rho_{31}, \\
   \frac{\partial}{\partial t} \rho_{31}
        = \frac{i}{2} \Omega_p +\frac{i}{2} \Omega_c \rho_{21}
        - \frac{\Gamma}{2} \rho_{31}, \\
    \frac{\partial}{\partial t} \rho_{21}
        = \frac{i}{2} (\Omega_c)^* \rho_{31} +\frac{i}{2}  (\Omega_s)^* \frac{\Omega'_c /\Gamma}{2\Delta /\Gamma_l} - \gamma \rho_{21},
\end{eqnarray}
where $\Omega'_c$, $\Delta$, and $\Gamma_l$ are the Rabi frequency, detuning, and spectral linewidth of the coupling's $|1\rangle \rightarrow |3\rangle$ transition, respectively.  Because $\Omega'_c$ and $\Omega_c$ are resulted from the same coupling field, they only differ by the ratio of the Clebsch-Gordan coefficients of the $|1\rangle \rightarrow |3\rangle$ and $|2\rangle \rightarrow |3\rangle$ transitions. In our case, $\Delta$ is about -1140$\Gamma$ and $\Gamma_l$ of cold atoms is approximately equal to the natural linewidth of the excited state $\Gamma$. Hence, such large $\Delta/ \Gamma_l$ makes the FWM effect negligible. Based on our experimental condition, the numerical predictions from Eqs.~(S1)-(S3) and from Eqs.~(S4)-(S7) are nearly the same. On the other hand, $\Gamma_l$ of hot or room-temperature atoms is about 100-fold larger due to the Doppler broadening, reducing $\Delta/ \Gamma_l$ by two orders of magnitude. Therefore, the FWM effect becomes rather significant in the hot medium of a large OD [24].

In the study of time-space-reversing method, the probe pulse was sent to the medium in the forward direction, stored by turning off $\Omega_c^+$, and retrieved in the backward direction by turning on $\Omega_c^-$. The equations are given by [31]
\begin{eqnarray}
    \frac{1}{c}\frac{\partial}{\partial t} \Omega^+_p
        + \frac{\partial}{\partial z} \Omega^+_p
        =  i\eta \rho^+_{31}, \\
    \frac{1}{c}\frac{\partial}{\partial t} \Omega^-_p
        - \frac{\partial}{\partial z} \Omega^-_p 
        =  i\eta \rho^-_{31}, \\
   \frac{\partial}{\partial t} \rho^+_{31}
        = \frac{i}{2} \Omega^+_p +\frac{i}{2} \Omega^+_c \rho_{21}
        - \frac{\Gamma}{2} \rho^+_{31}, \\
   \frac{\partial}{\partial t} \rho^-_{31}
       = \frac{i}{2} \Omega^-_p +\frac{i}{2} \Omega^-_c \rho_{21}
       - \frac{\Gamma}{2} \rho^-_{31}, \\
    \frac{\partial}{\partial t} \rho_{21}
        = \frac{i}{2} (\Omega^+_c)^* \rho^+_{31}
		+ \frac{i}{2} (\Omega^-_c)^* \rho^-_{31} - \gamma \rho_{21},
\end{eqnarray}
where the $+$ and $-$ signs indicate the fields and optical coherences in the forward and backward propagation directions, e.g. $\Omega_p^+$ is the probe pulse sent to the medium in the forward direction and $\Omega_p^-$ is that retrieved from the medium in the backward direction and, $\rho_{31}^+$ and $\rho_{31}^-$ are their corresponding optical coherences. 

In the paper, the theoretical predictions shown in Fig.~2(a) and the two blue lines in Fig.~2(b) were calculated with Eqs.~(S1)-(S3); those in Fig.~4(a) and the green line in Fig.~2(b) were with Eqs.~(S8)-(S12).
\end{document}